\documentclass[onecolumn,superscriptaddress,showpacs,nofootinbib,notitlepage]{aastex631}
\usepackage{graphicx,subfigure,amsmath,amsfonts,amssymb,multirow,mathrsfs,ulem,threeparttable}
\usepackage{hyperref}
\usepackage{epstopdf}
\usepackage{bm}
\begin{document}
\title{The detection prospect of the counter jet radiation in the late afterglow of GRB 170817A}

\correspondingauthor{Yi-Zhong Fan}
\email{yzfan@pmo.ac.cn}
\correspondingauthor{Yi-Ying Wang}
\email{wangyy@pmo.ac.cn}

\author{Jia-Ning Li}
\affiliation{Key Laboratory of Dark Matter and Space Astronomy, Purple Mountain Observatory, Chinese Academy of Sciences, Nanjing 210033, People's Republic of China}
\affiliation{School of Astronomy and Space Science, University of Science and Technology of China, Hefei, Anhui 230026, People's Republic of China}
\author[0000-0003-1215-6443]{Yi-Ying Wang}
\affiliation{Key Laboratory of Dark Matter and Space Astronomy, Purple Mountain Observatory, Chinese Academy of Sciences, Nanjing 210033, People's Republic of China}
\affiliation{School of Astronomy and Space Science, University of Science and Technology of China, Hefei, Anhui 230026, People's Republic of China}
\author[0000-0002-8385-7848]{Yun Wang}
\affiliation{Key Laboratory of Dark Matter and Space Astronomy, Purple Mountain Observatory, Chinese Academy of Sciences, Nanjing 210033, People's Republic of China}
\author[0000-0003-4977-9724]{Zhi-Ping Jin}
\affiliation{Key Laboratory of Dark Matter and Space Astronomy, Purple Mountain Observatory, Chinese Academy of Sciences, Nanjing 210033, People's Republic of China}
\affiliation{School of Astronomy and Space Science, University of Science and Technology of China, Hefei, Anhui 230026, People's Republic of China}
\author[0000-0001-9078-5507]{Stefano Covino}
\affiliation{INAF/Brera Astronomical Observatory, via Bianchi 46, I-23807 Merate (LC), Italy}
\author[0000-0002-8966-6911]{Yi-Zhong Fan}
\affiliation{Key Laboratory of Dark Matter and Space Astronomy, Purple Mountain Observatory, Chinese Academy of Sciences, Nanjing 210033, People's Republic of China}
\affiliation{School of Astronomy and Space Science, University of Science and Technology of China, Hefei, Anhui 230026, People's Republic of China}

\newcommand{\ud}{\mathrm{d}}
\begin{abstract}
The central engine of a Gamma-Ray Burst (GRB) is widely believed to launch a pair of oppositely moving jets, i.e. the forward jet moving towards us and the counter jet regressing away. The forward jet generates the radiation typically observed in GRBs, while the counter jet has not been detected yet due to its dimness. 
GRB 170817A, a short burst associated with a binary neutron star merger event, is a nearby event ($z=0.0097$) with an off-axis structured energetic forward jet and hence probably the most suitable target for searching the counter jet radiation.  
Assuming the same properties for the forward and counter jet components as well as the shock parameters, the fit to the multi-wavelength afterglow emission of GRB 170817A suggests a peak time $\sim {\rm quite~a~few}\times 10^{3}$ day of the counter jet radiation, but the detection prospect of this new component is not promising. 
Anyhow, if the shock parameters ($\epsilon_{\rm e}$ and $\epsilon_{\rm B}$) of the counter jet component are (a few times) higher than that of the forward shock e.g., the posterior results of the magnetic energy fraction of the forward shock and the counter jet are  $\log_{10}\epsilon_{\rm B}=-4.23^{+1.42}_{-0.69}$ and $\log_{10}\epsilon_{\rm B, cj}=-3.65^{+3.06}_{-2.11}$, respectively), as still allowed by the current data, 
the counter jet afterglow emission will be enhanced and hence may be detected. 
A few hour exposure by JWST in F356W band will stringently test such a scenario.
\end{abstract}

\section{Introduction}\label{sec:1}
At 12:41:04 UTC on August 17, 2017, the first binary neutron star (BNS) merger GW170817 was observed by the advanced LIGO and Advanced Virgo detectors. Intriguingly, just $\sim 1.4$ seconds after the merger, the Fermi Gamma-Ray Burst Monitor detected a low-luminosity short Gamma-Ray Burst (SGRB), called GRB 170817A, which was subsequently confirmed to be the electromagnetic counterpart of GW170817 \citep[e.g.,][]{2017ApJ...848L..12A, 2017ApJ...848L..14G}. The kilonova emission dominates the observed signal in ultraviolet, optical and infrared bands for a few weeks \citep{2017ApJ...848L..17C,2017Sci...358.1565E,2017Natur.551...67P,2017Natur.551...75S}
The synchrotron afterglow was first observed in X-ray at 9 days after the merger \citep{2017Natur.551...71T}, followed by detections in radio and optical at 16 and 110 days post-merger \citep{2017Sci...358.1579H, 2018MNRAS.481.2581L, 2018A&A...613L...1D, 2018A&A...619A..18S, 2019A&A...628A..18S, 2019Sci...363..968G}, respectively.
The evolution of multi-band afterglow is marked by a steady shallow rise in the initial few tens of days, reaching a peak at approximately 164 days, and then gradually declining thereafter \citep{2018ApJ...863L..18A, 2018ApJ...858L..15D, 2018ApJ...868L..11M, 2018ApJ...855..103P, 2018MNRAS.478L..18T, 2019MNRAS.489.1919T, 2019ApJ...870L..15L, 2021ApJ...922..154M}. 
In addition, GRB 170817A was a low luminosity event \citep{2018ApJ...856L..18M}, and its isotropic equivalent energy is significantly lower than typical SGRBs \citep{2014ARA&A..52...43B, 2015ApJ...815..102F, 2017ApJ...848L..13A, 2017ApJ...848L..14G}. 
These phenomena undoubtedly indicate that GRB 170817A is not a typical on-axis SGRB, instead it is generated by a highly off-axis structured ejecta rather than uniform top-hat ejecta \citep{2017ApJ...848L..21A, 2017ApJ...848L..25H, 2017Sci...358.1559K, 2018MNRAS.473L.121K, 2019ApJ...876L..28D}. Note that off-axis highly-structured ejecta can also effectively enhance the GRB/GW association rate, as pointed out by \citet{2018ApJ...857..128J} initially before the release of the data of GW170817.

It is challenging to fully resolve the structure of the ejecta of GRB 170817A. Nevertheless, the off-axis uniform jet model is disfavored as it fails to provide a satisfactory explanation for the early X-ray and radio data \citep{2017Natur.551...71T, 2017Sci...358.1579H, 2018ApJ...857..128J, 2021Univ....7..329L}. Two alternative scenarios have attracted wide attention, including a structured jet with an angular profile of Lorentz factors and energy \citep{2018MNRAS.478L..18T, 2018MNRAS.473L.121K,2021ApJ...908..200W} and a mildly-relativistic isotropic cocoon \citep{2017Sci...358.1559K, 2018MNRAS.478L..18T}. The late time temporal behavior of the afterglow emission of GRB 170817A, however, challenges most models of choked jet/cocoon systems, and instead supports the emergence of a relativistic structured jet \citep{2018ApJ...868L..11M, 2019MNRAS.489.1919T}. For long duration GRBs, structured jets have been proposed, e.g., by \cite{2002MNRAS.332..945R, 2002ApJ...571..876Z, 2004ApJ...606L..37N, 2017MNRAS.472.4953L}. Such a possibility has also been considered for short GRB afterglow modeling firstly by \citet{2007ApJ...656L..57J}.  In a specific scenario, the energy and bulk Lorentz factor in the source frame exhibit Gaussian-like angular variations \citep{2017MNRAS.472.4953L,2018ApJ...857..128J}, which is found to be in line with the observed characteristics of GRB~170817A  \citep{2018MNRAS.478..733L, 2018ApJ...867...57R, 2019MNRAS.485.1435H, 2019ApJ...870L..15L}. Therefore, in this work we adopt the off-axis Gaussian jet model to reproduce the properties of the afterglow emission of GRB 170817A. 

Considering the huge energies and short time scales of GRBs, the powerful prompt emission should originate from highly relativistic jets \citep{1992Natur.357..472U, 1992ApJ...394L..33B, 1999ApJ...518..356P, 2000PhR...325...83L}. Due to the symmetry of the central engine system, the jets are expected to be dual-sided \citep{2004ApJ...614L..17L, 2004MNRAS.354...86R}, with one branch, known as the forward jet, moving towards the observer, and the other, known as the counter jet, regressing away. The shock wave driven by the jet moving into the surrounding medium generates a long-lasting broadband synchrotron afterglow \citep{1997ApJ...476..232M, 1998ApJ...497L..17S}. Initially, the radiation from the forward jet typically dominates in intensity and observable flux, while the simultaneous emission from the counter jet is highly suppressed due to its relativistic motion, resulting in negligible contribution to the observed afterglow emission. However, as the core of the jets get decelerated to be sub-relativistic or non-relativistic \citep{2004ApJ...614L..17L} in the later stages, counter jet influence on the observed flux becomes more pronounced. Due to the delay in light propagation time, the contribution of the counter jet is superimposed on the emission of the decaying late forward jet, which may result in a short plateau or re-brightening in the late-time afterglow light curve \citep{2003ApJ...593L..81G, 2004ApJ...614L..17L, 2009ApJ...698.1261Z, 2009A&A...505.1213W, 2010SCPMA..53S.259W}. Unless the source is nearby, the detection of such a signature is challenging and it has not yet been detected so far \citep{2010AIPC.1279..460W, 2018PTEP.2018c3E01Y}. GRB 170817A provides us with an unprecedented chance. A relativistic moving jet was confirmed robustly from the Very Long Baseline Interferometry (VLBI) measurements and the turnover in radio observations by \citet{2018ApJ...863L..18A}. \citet{2018MNRAS.478.4128G} proposed that the counter jet would start to impact the afterglow light curve of GRB 170817A beyond 1000 days, resulting in a flattened light curve. \citet{2019ApJ...870L..15L} instead suggested that the contributions of the counter jet would not become evident in the light curve until $\sim 10^4$ days. \citet{2019ApJ...880...39L} modeled the multi-band afterglow of GRB 170817A and found out the appearance of the counter-jet at nearly 2500 days post-merger in the radio afterglow emission. 
In this work we re-visit this issue with the more complete afterglow data set of GRB 170817A. 
We adopt a Gaussian structured jet model to calculate the forward and counter jet components in the afterglow emissions of GRB 170817A and evaluate whether the contribution of the counter jet to the late-time afterglow emission is discernible. We also investigated the sensitivity of the current GRB Afterglow Detector Network to ascertain the detectability of the counter jet. 

The work is arranged as the follows. In Sec.~\ref{sec:2}, we outline the structured jet model considered in this paper. In Sec.~\ref{sec:3}, we detail the methods employed, the observed data, and the Bayesian inference results. In Sec.~\ref{sec:4}, we discuss the emission of the counter jet. In the end, the Sec.~\ref{sec:5} shows the conclusion. 

\section{Model}\label{sec:2}
If the ejecta of a GRB is structured, then for the observer the isotropic-equivalent energy of the shock wave is closely related to the angle from the jet axis $\theta$.
The energy of a structured jet varies with angle due to the jet-launching mechanism \citep{2003ApJ...594L..23V, 2003ApJ...584..937V} or the interaction of the ejecta with the surrounding matter during its breakout \citep{2002bjgr.conf..146L, 2003ApJ...594L..19L, 2003ApJ...586..356Z, 2004ApJ...608..365Z, 2005A&A...436..273A, 2005ApJ...629..903L, 2010ApJ...723..267M, 2015MNRAS.447.1911P}. For GRB 170817A, we assume that the dual-sided ultra-relativistic jets, comprising a forward jet and a counter jet, were launched from the central engine and have identical characteristics, including the semi-opening angle, initial Lorentz factor, and energy. As such, the properties of the counter jet component can be reliably constrained because most of or all parameters of the counter jet are assumed to be the same as the forward ones, and the observed afterglow emission, in particular at early times, are dominated by the forward jet.
 
Our treatments are mainly following \cite{2020ApJ...896..166R} with some adjustments and extensions. In general, the energy of the Gaussian structured jet varies with the angle as 
\begin{equation}\label{gaussian_energy}
   E(\theta)= \begin{cases}
        E_{0}~{\rm exp} \bigl(-\frac{\theta^2}{2\theta_{\rm c}^2}\bigr) & \mathrm{\theta \leq \theta_{\rm w}} \\
       0 & \mathrm{\theta > \theta_{\rm w}} 
   \end{cases},
\end{equation}
where 
$\theta_{\rm c}$ determines the width of the jet core,  and $\theta_{\rm w}$ represents the truncation angle of the Gaussian wings. Beyond the truncation angle, there is no ejecta at all. The Lorentz factor is assumed to follow an analogous angular dependence.

Assuming the number density of the ambient medium is $n_{0}$,  the number density of the shocked medium is $n'\approx 4n_{0}\Gamma$, where $\Gamma$ is the Lorentz factor of the shocked medium. And the thermal energy of the shocked medium (i.e., the fluid behind the shock) is $e_{\rm th} = (\Gamma-1)n'm_{\rm p}c^2$, where $m_{\rm p}$ is the proton mass and $c$ represents the speed of light in the vacuum \citep{1998ApJ...497L..17S}. Therefore, the magnetic field strength is $B'=\sqrt{4\pi\epsilon_{\rm B} e_{\rm th}}$, where $\epsilon_{\rm B}$ is the fraction of the thermal energy given to the magnetic field \citep{2020ApJ...896..166R}.
The shock-accelerated electrons are assumed to follow a power-law distribution with an index of $-p$ ($p>2$), with a minimum Lorentz factor of $\gamma_{\rm m}\approx (\epsilon_{\rm e}/\xi_{\rm N})[(p-2)/(p-1)](\Gamma-1)[m_{\rm p}/m_{\rm e}]$, where $m_{\rm e}$ is the electron mass and $\xi_{\rm N}$ is the fraction of accelerated electrons ($\xi_{\rm N}=1$ is commonly assumed). The cooling Lorentz factor of the electrons can be estimated as $\gamma_{\rm c}=6\pi m_{\rm e}c/(\sigma_{\rm T}\Gamma {B'}^{2}t)$, where $\sigma_{\rm T}$ is the Thompson cross section and $t$  is the time in the frame of the observer \citep{1998ApJ...497L..17S}. The spectral energy distribution can be described with the minimum frequency $\nu'_{\rm m}= 3\gamma_{\rm m}^2 eB'/(4\pi m_{\rm e}c)$, the cooling frequency $\nu'_{\rm c}=3 \gamma_{\rm c}^2 eB'/(4\pi m_{\rm e}c)$, and the self-absorption frequency $\nu'_{\rm a}$ \citep{1979rpa..book.....R}, where $e$ is the charge of the electron and the superscript ``$'$" represents the parameter measured in the rest-frame of the fluid. In the case of $\nu'_{\rm a}$ is below the observer's frequency $(\nu')$ as well as $\min\{\nu'_{\rm m},\nu'_{\rm c} \}$, the rest-frame emissivity can be described by \citet{2021ApJ...908..200W} as
\begin{equation}
    \epsilon'_{\nu'} = \epsilon_{\rm p}\times\begin{cases}
        (\nu'/\nu'_{\rm m})^{1/3} & \nu'<\nu'_{\rm m}<\nu'_{\rm c}, \\
        (\nu'/\nu'_{\rm m})^{-(p-1)/2} & \nu'_{\rm m}<\nu'<\nu'_{\rm c}, \\
        (\nu'_{\rm c}/\nu'_{\rm m})^{-(p-1)/2}(\nu'/\nu'_{\rm c})^{-p/2} & \nu'_{\rm m}<\nu'_{\rm c}<\nu'; \\
        (\nu'/\nu'_{\rm c})^{1/3} & \nu'<\nu'_{\rm c}<\nu'_{\rm m}, \\
        (\nu'/\nu'_{\rm c})^{-1/2} & \nu'_{\rm c}<\nu'<\nu'_{\rm m}, \\
        (\nu'_{\rm m}/\nu'_{\rm c})^{-1/2}(\nu'/\nu'_{\rm m})^{-p/2} & \nu'_{\rm c}<\nu'_{\rm m}<\nu',
    \end{cases}
\end{equation}
where $\epsilon_{\rm p}$ represents the peak emissivity of the synchrotron spectrum, which is  given as \citep{2020ApJ...896..166R}
\begin{equation}
    \epsilon_{\rm p}=\frac{p-1}{2}\frac{\sqrt{3}e^{3}\xi_{\rm N}n'B'}{m_{\rm e}c^{2}}.
\end{equation}

At the rest frame time $T$, 
the forward shock radius is $R(T; \theta)$.
The photon emitted from blast wave at time $T$ would be observed by the observer at time $t$, i.e.,
the observation time can be expressed as
\begin{equation}
    t=T-\frac{\mu}{c}R(T; \theta),
\end{equation}
where $\mu =\cos(\theta_{\rm v}-\theta)$ and $\theta_{\rm v}$ is the observer's viewing angle.

Finally, the observed flux at the time $t$ and frequency $\nu_{\rm v}=\delta \nu'$ ($\delta=\gamma^{-1}(1-\beta \mu)^{-1}$ represents the Doppler factor of the emitting fluid for the observer) is given as
\citep{2010ApJ...722..235V,2020ApJ...896..166R}
\begin{equation}\label{flux1}
	F_{\rm v}(t,\nu_{\rm v})=\frac{1+z}{4\pi d_{\rm L}^2}\int d\Omega R^2\Delta R\delta^2\epsilon'_{\nu'},
\end{equation}
where $z$ is the redshift of the source, and $d_{\rm L}$ is the luminosity distance.

Assuming the angle between the observer's line of sight and the forward jet axis as $\theta_{\rm v,fj}$, it turns out that $\theta_{\rm v,fj} \leq \pi /2$. Consequently, the angle between the observer's line of sight and the counter jet axis can be denoted as $\theta_{\rm v,cj}= \pi - \theta_{\rm v,fj}$, and is greater than $\pi /2$. Note that the subscripts "$\rm fj$" and "$\rm cj$" represent the forward jet and counter jet, respectively. Therefore, the observation angle affects the observation time and the flux calculation. To simplify the calculation, it is assumed that the observed radiation flux is a simple superposition of two jet components. 
The observed total flux at the observed time $t$ and frequency $\nu_{\rm v}$ is
\begin{equation}
    F_{\rm v,tot}(t,\nu_{\rm v})=F_{\rm v,fj}(t_{\rm fj},\nu_{\rm v}; \theta_{\rm fj})+F_{\rm v,cj}(t_{\rm cj},\nu_{\rm v}; \theta_{\rm v,cj}).
\end{equation}
In this work, we employ an off-axis Gaussian structured jet model featuring both forward and counter jet branches to numerically calculate the afterglow emission of GRB 170817A. The details are presented in Section.~{\ref{3.2}}.

\section{Method and Result}\label{sec:3}
\subsection{Method}\label{3.1}
The current observations suggest that the afterglow emission of GW170817$/$GRB170817A appears to be consistent both with the decelerating relativistic gamma-ray burst outflow and the onset of kilonova afterglow \citep{2023ApJ...943...13W}, which results from the sub-relativistic combined interaction with the surrounding medium. The latter process is similar to the production of afterglow by cocoons.
Thus, the kilonova afterglow can be approximated to the spherical cocoon afterglow in the subrelativistic range \citep{2023arXiv231206286H}. Furthermore, it is crucial to consider the lateral spreading at late times when determining the various physical properties of GRBs \citep{2023arXiv231109297G}, since the lateral spreading will generate shallower-decaying light curves which are more consistent with the real observations \citep{2023ApJ...943...13W}. Therefore, in our modeling, we take into account the above issues.
For our purpose, we take the open source package {\tt Afterglowpy} \citep{2020ApJ...896..166R} to calculate the afterglow light curves.

To constrain the jet profile of GRB 170817A, we utilize the dataset from \citet {2022ApJ...938...12B} and the latest detection \citep{2022GCN.32065....1O} in the X-ray band.  
The whole dataset covers a time interval ranging from 9.2 to 1,674 days in radio, optical, and X-ray bands, respectively.
In our model, the total afterglow emission flux can be written as
\begin{equation}\label{eq:7}
    F_{\rm v,tot}=F_{\rm v,fj}+F_{\rm v,cj}+F_{\rm v,KN},
\end{equation}
where the subscript $\rm KN$ represents the kilonova afterglow component.
All of the free parameters hidden in Equation~(\ref{eq:7}) are listed in Table.~\ref{tb:prior}. 
Note that, the fraction of accelerated electrons $\xi_{\rm N}$ is fixed to 1, and the luminosity distance $d_{\rm L}$ is set to $1.23 \times 10^{26} \rm {cm}$ \citep{2020ApJ...896..166R}. For the kilonova afterglow, the thermal energy fraction of electrons $\epsilon_{\rm e}$, the thermal energy fraction in the magnetic field $\epsilon_{\rm B}$, and the electron index $p$ are independent of those of the GRB ejecta. The energy distribution of the kilonova outflow is assumed to be $\propto U^{-k}$ for $U_{\rm min}\leq U \leq U_{\rm max}$, where $U$ is the 4-velocity. Following \citet{2017ApJ...850L..37P}, \citet{2020ApJ...891..152H} and \citet{2024ApJ...961....9F}, $U_{\rm max}$ is taken to be within $[0.15,0.45]$ to ensure that the highest velocity is smaller than $0.4c$.

These free parameters can be constrained by fitting the synthetic multi-wavelength light curves. 
In the Bayesian statistical framework, the likelihood function can be written as 
\begin{equation}
    \mathcal{L}(F_{\rm obs} \mid \boldsymbol{\hat{\theta}})=\prod^{N}_{i} \frac{1}{\sqrt{2\pi}\sigma_i} {\rm exp} \biggl[ -\frac{1}{2} \biggl(\frac{f_{\rm obs}(x_i)-y_i}{\sigma_i} \biggr)^2 \biggr],
\end{equation}
where $f_{\rm obs}(x_i)$ and $\sigma_i$ represent the observed afterglow light curve data and their uncertainties, respectively; $y_i$ is the predicted value of the afterglow model at $x_i$. The posterior probability density is 
\begin{equation}
    P(\boldsymbol{\hat{\theta}} \mid F_{\rm obs})\propto \mathcal{L}(F_{\rm obs} \mid \boldsymbol{\hat{\theta}}) P(\boldsymbol{\hat{\theta}}),
\end{equation}
where $\mathcal{L}(F_{\rm obs} \mid \boldsymbol{\hat{\theta}})$ is the likelihood function, $P(\boldsymbol{\hat{\theta}})$ is the prior distribution.
We use {\tt Nessai} as a nested sampler, which is in combination with {\tt Bilby}, for Bayesian statistical inference and parameter estimation.
Furthermore, the superluminal motion of the jet observed with VLBI gives an independent constraint of $0.2< \theta_{\rm v}(\frac{d_{\rm L}}{41 Mpc})<0.5$ \citep{2018Natur.561..355M}, providing a tighter limitation for $\theta_{\rm v}$ 
and therefore increasing the estimated precision obviously \citep{2019NatAs...3..940H}.
As shown in Table~\ref{tb:prior}, the prior distributions of these parameters are consistent with \cite{2023ApJ...943...13W} and \cite{2023arXiv231002328R}. 

\begin{table*}[ht!]
\begin{ruledtabular}
\centering
\caption{Prior Distributions and Posterior Results of the parameters for GRB 170817A and kilonova afterglow}
\label{tb:prior}
\begin{tabular}{llccc}
\multirow{2}{*}{Names} &\multirow{2}{*}{Parameters}   &Priors of Parameter    &\multicolumn{2}{c}{Posterior Results\textsuperscript{c}}  \\
& &Inference  &Model (A) &Model (B) \\\hline            
Viewing angle &$\theta_{\rm v}$($\rm rad$)           &Sine(0, $\pi$)                   &$0.48^{+0.03}_{-0.08}$                 &$0.48^{+0.03}_{-0.08}$ \\
Half opening angle\textsuperscript{a} &$\theta_{\rm c}$($\rm rad$)           &Uniform(0, $\pi/2$)              &$0.07^{+0.01}_{-0.01}$                 &$0.07^{+0.01}_{-0.01}$ \\
Outer truncation angle\textsuperscript{a} &$\theta_{\rm w}$($\rm rad$)           &Uniform(0, $\pi/2$)              &$0.55^{+0.31}_{-0.27}$                 &$0.56^{+0.30}_{-0.27}$ \\ 
On-axis energy &$\log_{10} E_{0}$($\rm erg$)           &Uniform(45, 57)                  &$54.57^{+0.81}_{-1.35}$                 &$54.41^{+0.88}_{-1.46}$ \\
Circumburst density &$\log_{10} n_{0}$($\rm cm^{-3}$)       &Uniform(-6, 0)                   &$-0.57^{+0.51}_{-1.31}$                 &$-0.72^{+0.65}_{-1.40}$ \\
Spectral index &$p$                   &Uniform(2, 2.5)                  &$2.12^{+0.02}_{-0.01}$                 &$2.12^{+0.02}_{-0.01}$ \\	
Electron energy fraction &$\log_{10} \epsilon_{\rm e}$           &Uniform(-6, 0)                   &$-3.73^{+1.21}_{-0.75}$                 &$-3.58^{+1.31}_{-0.80}$ \\
Magnetic energy fraction &$\log_{10} \epsilon_{\rm B}$          &Uniform(-6, 0)                   &$-4.39^{+1.34}_{-0.57}$                 &$-4.23^{+1.42}_{-0.69}$ \\   
Electron energy fraction &\multirow{2}{*}{$\log_{10} \epsilon_{\rm e,cj}$}            &\multirow{2}{*}{Uniform(-6, 0)}                   &\multirow{2}{*}{$\dots$}                 &\multirow{2}{*}{$-4.14^{+2.75}_{-1.66}$} \\
\quad of counter jet\textsuperscript{b} & & & &\\
Magnetic energy fraction &\multirow{2}{*}{$\log_{10} \epsilon_{\rm B,cj}$}             &\multirow{2}{*}{Uniform(-6, 0)}                   &\multirow{2}{*}{$\dots$}                &\multirow{2}{*}{$-3.65^{+3.06}_{-2.11}$} \\
\quad of counter jet\textsuperscript{b} & & & &\\\hline
Maximum 4-velocity of  &\multirow{2}{*}{$U_{\rm max}$}            &\multirow{2}{*}{Uniform(0.15, 0.45)}               &\multirow{2}{*}{$0.36^{+0.08}_{-0.15}$}                 &\multirow{2}{*}{$0.36^{+0.08}_{-0.15}$} \\
\quad outflow & & & & \\
Minimum 4-velocity of &\multirow{2}{*}{$U_{\rm min}$}          &\multirow{2}{*}{Uniform(0.1, 0.15)}               &\multirow{2}{*}{$0.11^{+0.03}_{-0.06}$}                 &\multirow{2}{*}{$0.11^{+0.03}_{-0.05}$} \\
\quad outflow & & & &\\
Normalization of outflow's  &\multirow{2}{*}{$\log_{10} E_{\rm i}$($\rm erg$)}           &\multirow{2}{*}{Uniform(45, 50)}       &\multirow{2}{*}{$48.12^{+1.48}_{-2.28}$}                 &\multirow{2}{*}{$48.07^{+1.49}_{-2.26}$} \\
\quad energy distribution & & & &\\
Power-law index of the energy &\multirow{2}{*}{$k$}             &\multirow{2}{*}{Uniform(0.5, 4)}                   &\multirow{2}{*}{$1.47^{+2.08}_{-0.90}$}                 &\multirow{2}{*}{$1.50^{+2.07}_{-0.93}$} \\
\quad velocity distribution & & & &\\
Spectral index &$p_{\rm KN}$            &Uniform(2.0, 2.5)                &$2.11^{+0.09}_{-0.08}$                 &$2.11^{+0.10}_{-0.08}$ \\
Electron energy fraction &$\log_{10} \epsilon_{\rm e,KN}$             &Uniform(-5, 0)                   &$-0.46^{+0.42}_{-1.10}$                 &$-0.47^{+0.43}_{-1.24}$ \\
Magnetic energy fraction &$\log_{10} \epsilon_{\rm B,KN}$             &Uniform(-5, 0)                   &$-3.04^{+1.52}_{-1.51}$                 &$-2.91^{+1.62}_{-1.60}$ \\
Fraction of electrons that &\multirow{2}{*}{$\xi_{\rm N,KN}$}             &\multirow{2}{*}{Uniform(0, 1)}   &\multirow{2}{*}{$0.48^{+0.46}_{-0.43}$}   &\multirow{2}{*}{$0.47^{+0.47}_{-0.43}$} \\ 
\quad get accelerated & & & &\\
  
\end{tabular}
\begin{tablenotes}
  \item[a] \textsuperscript{a} $\theta_{\rm c}$ and $\theta_{\rm w}$ are limited to $0<\theta_{\rm w}/ \theta_{\rm c}<12$.
  \item[b] \textsuperscript{b} These two parameters are only used when the parameters of the forward jet and the counter jet are different.
  \item[c] \textsuperscript{c} The posterior results for each model are at the $68.3\%$ credible level.
  \\
\end{tablenotes}
\end{ruledtabular}
\end{table*}

\subsection{Result}\label{3.2}
\begin{figure}[ht!]
	\centering	
    \subfigure[]{
    \includegraphics[width=0.48\textwidth]{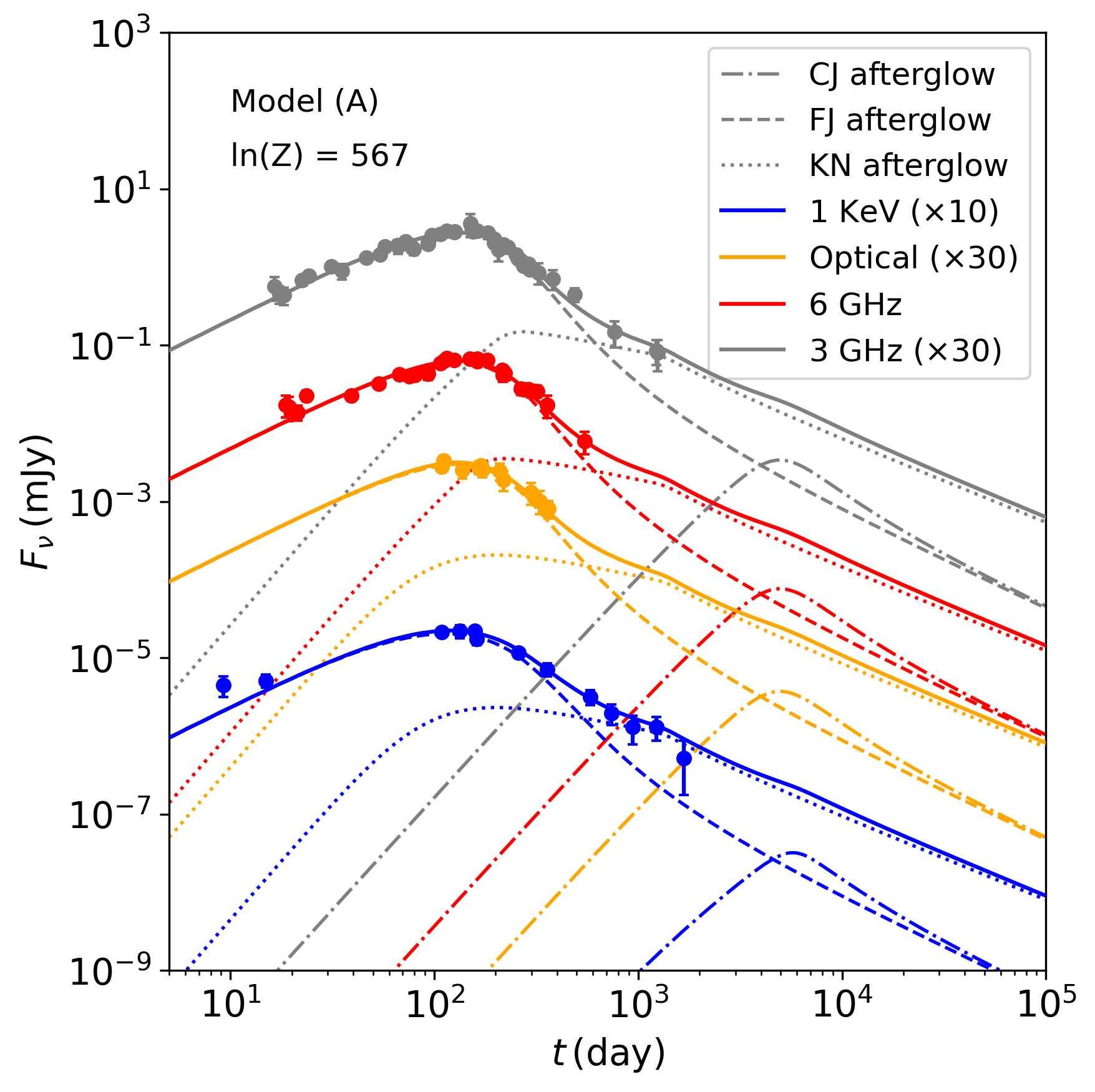}
    \label{subfig:1a}}
    \subfigure[]{
    \includegraphics[width=0.48\textwidth]{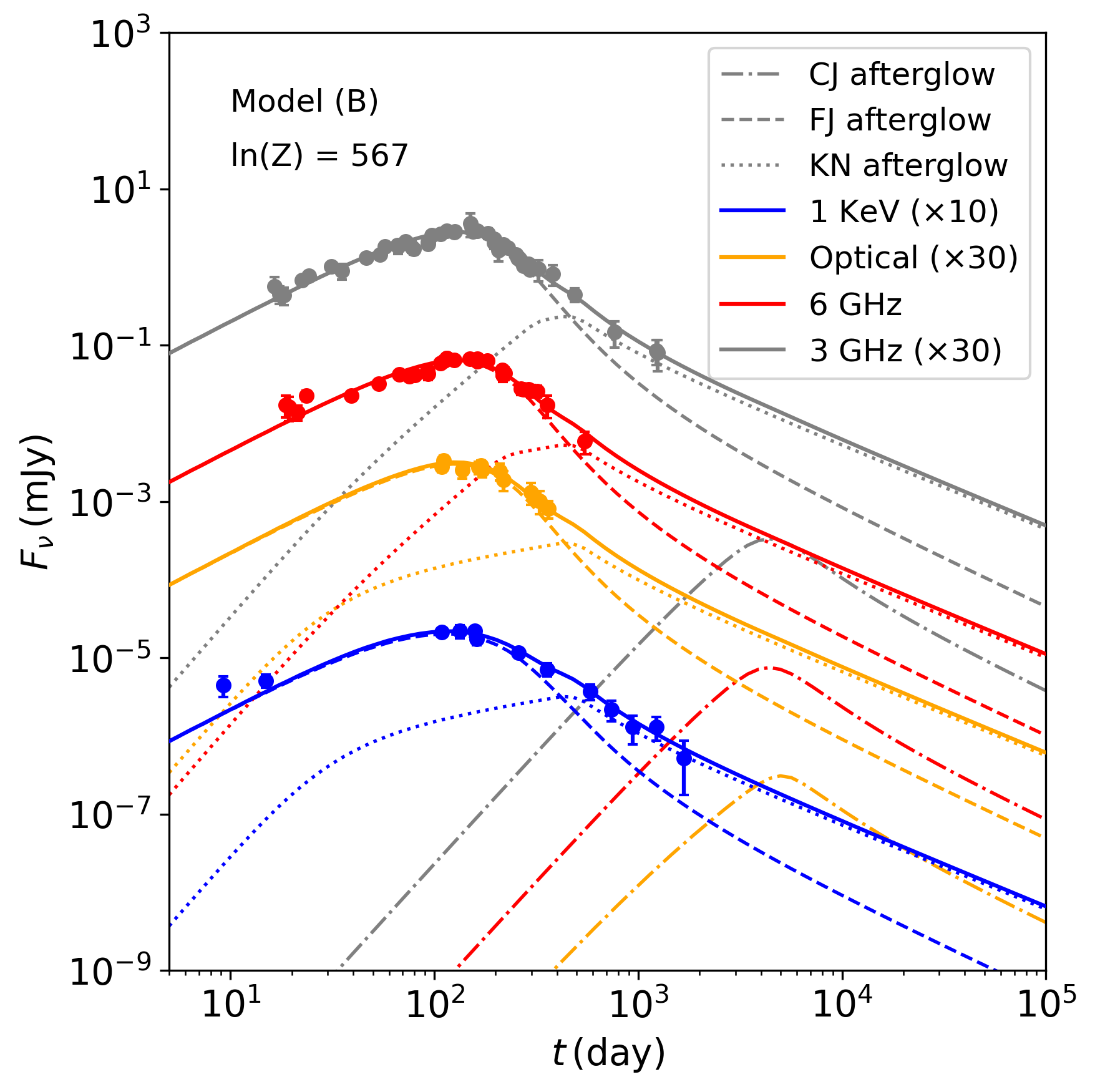}
    \label{subfig:1b}}
    \subfigure[]{
    \includegraphics[width=0.48\textwidth]{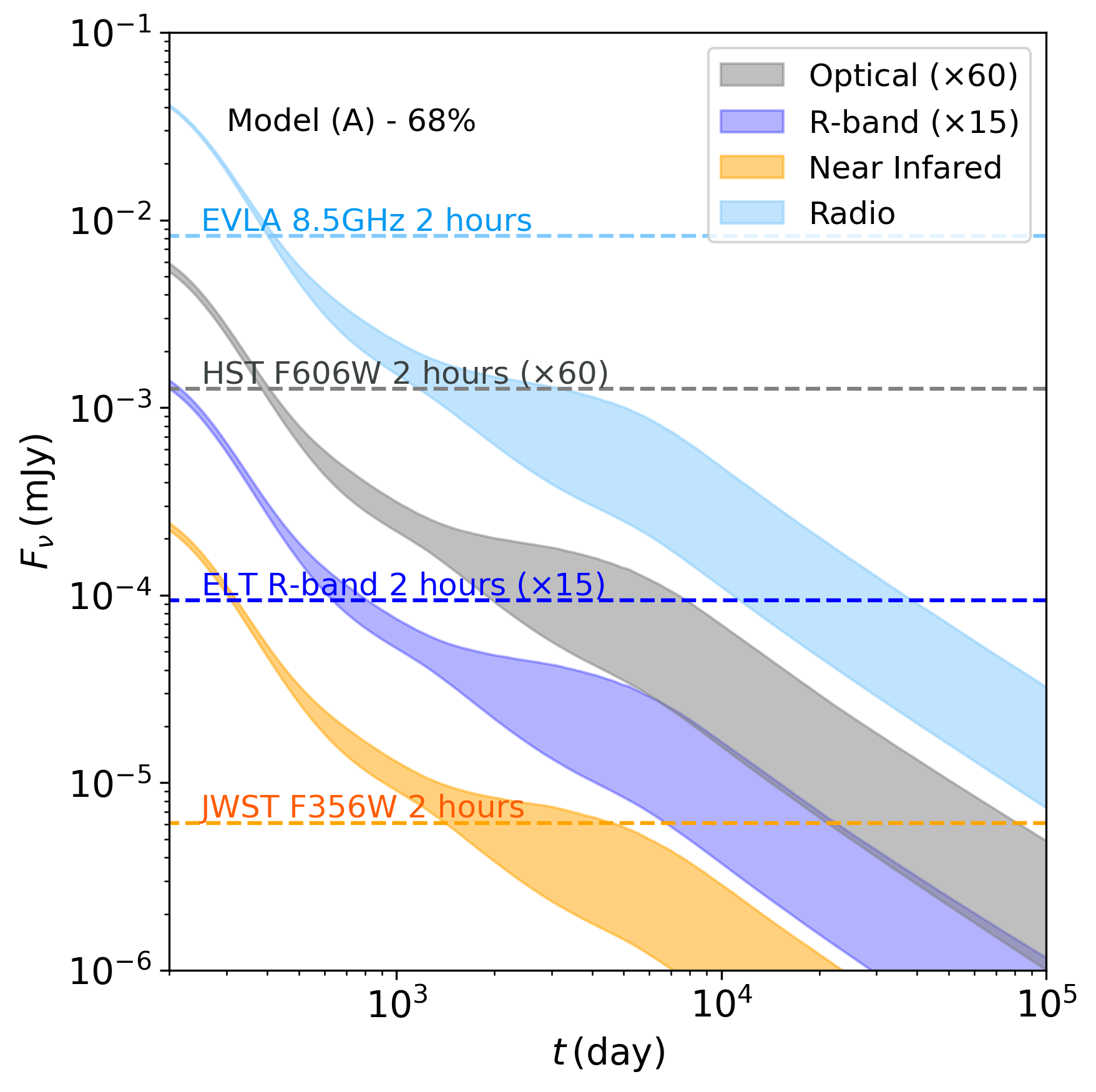}
    \label{subfig:1c}}
    \subfigure[]{
    \includegraphics[width=0.48\textwidth]{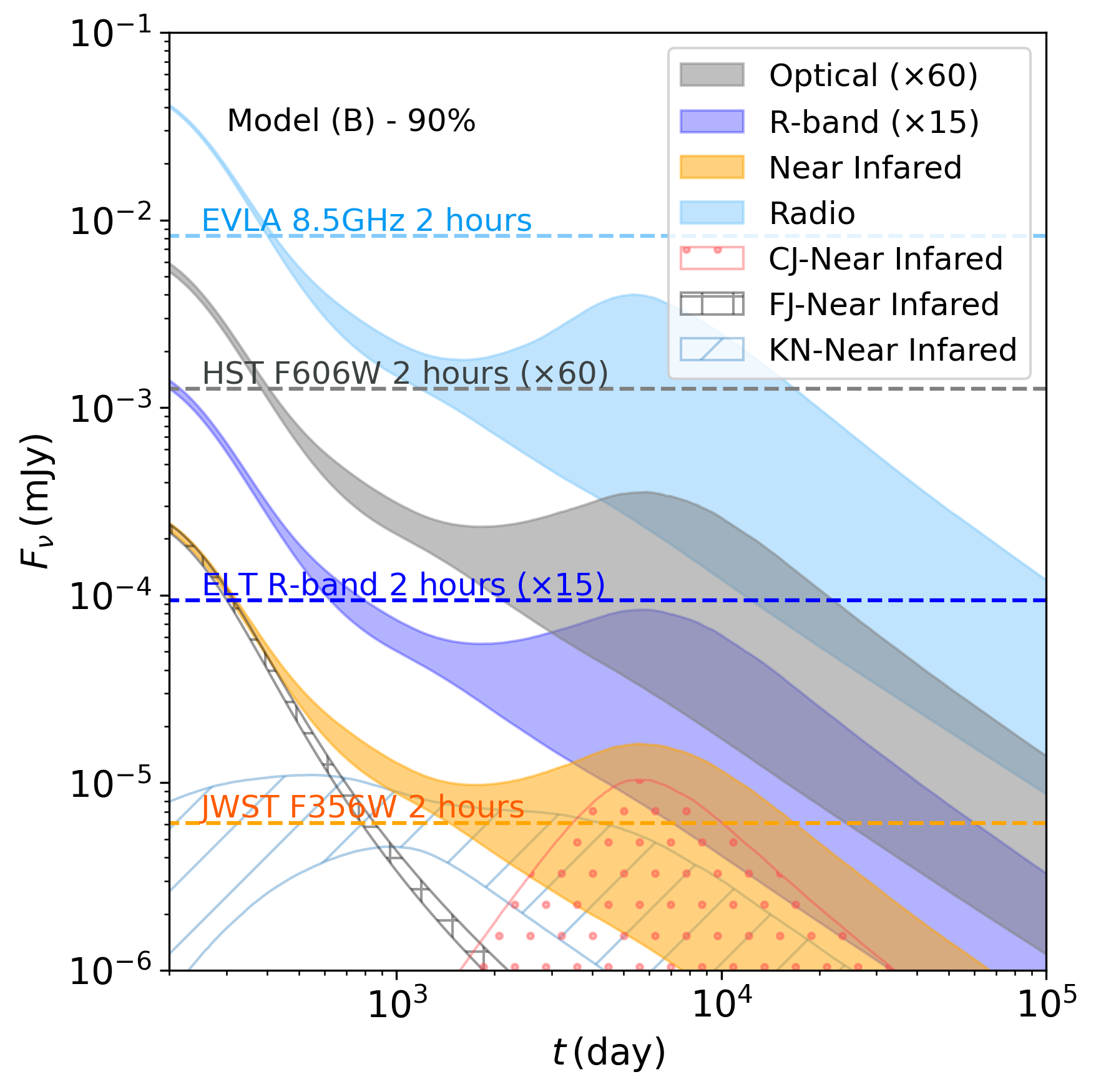}
    \label{subfig:1d}}
	\caption{\small The best fitting light curves for {\it Model (A)} and {\it Model (B)} are shown in panel (a) and (b), respectively. The observation data points and the theoretical light curves at 3 GHz, 6 GHz, $5.06 \times 10^{14}$ Hz (Optical) and 1 KeV bands are shown in grey, red, orange, and blue, respectively. Each band's flux is rescaled for a better view. 
    In each panel, the solid lines represent the best fitting of the afterglow, including the contributions of the forward jet, counter jet, and kilonova afterglow components. The dashed lines represent the forward jet afterglow component, the dotted lines represent the kilonova afterglow and the dot-dashed lines represent the counter jet afterglow.
    In {\it Model (A)}, the peak flux of the counter jet radiation is  substantially lower than the simultaneous kilonova radiation, which is hard to be detected.
    In {\it Model (B)}, the counter jet has a slightly higher peak radiation.
    The detectability of the counter jet afterglow by HST in F606W (gray), JWST in F356W (orange), ELT in R band (blue) and EVLA (sky blue) with 2 hours of exposure are shown in panel (c) and (d), respectively. In F356W band, the components of the forward jet, the counter jet, and the kilonova are presented in pink dotted region, blue slash region, and the grey crosswise region, respectively.
    Different from panels (a) and (b), now the $68 \%$ credible regions of the light curve are presented. The detection prospect in the case of $Model (B)$ may be promising in JWST/F356 band but challenging for HST/F606W and EVLA.}
	\label{fig:bestfit}
\end{figure}

\begin{figure*}[ht!]
	\centering
	\includegraphics[width=0.99\textwidth]{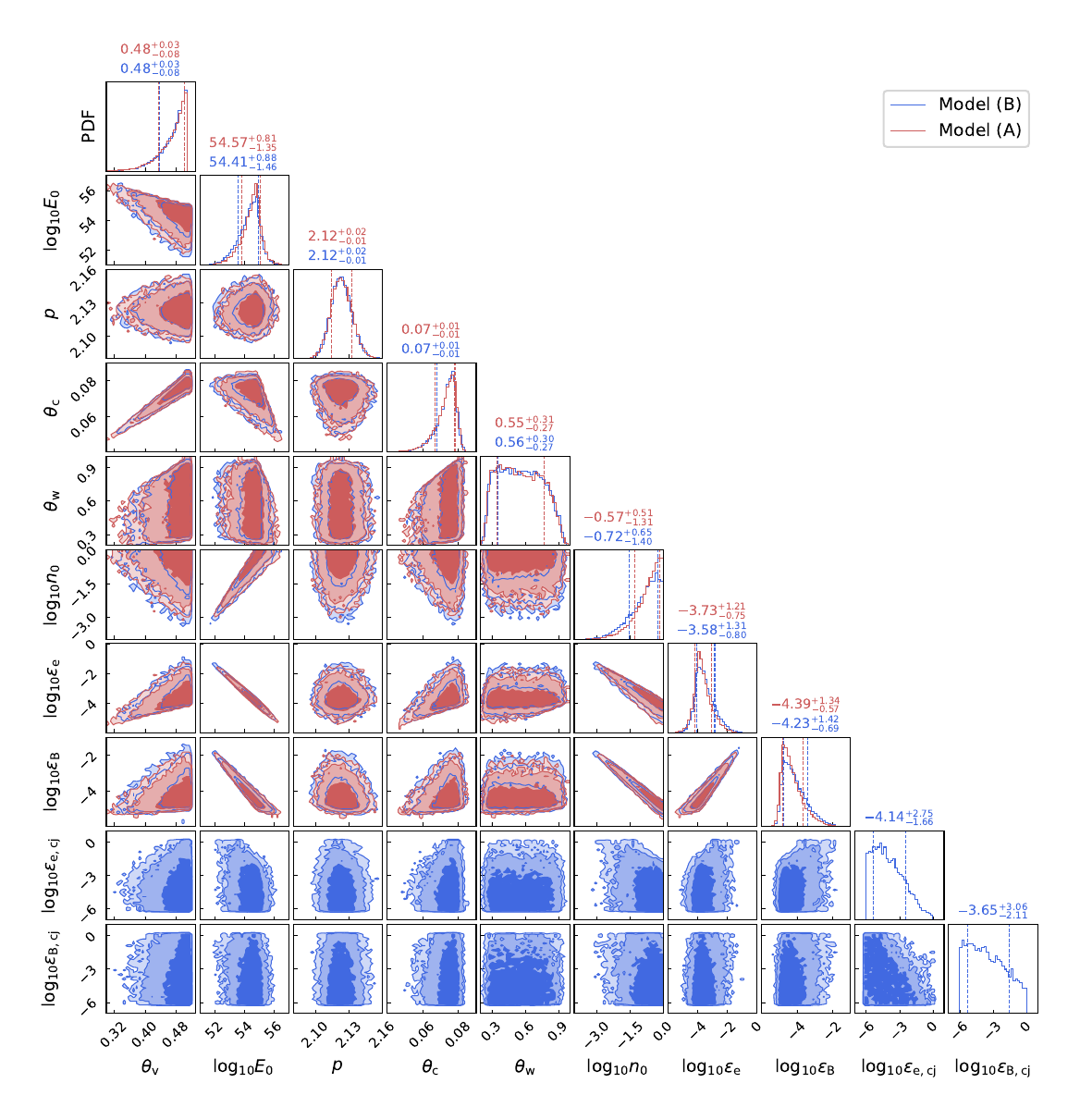}
    \caption{\small The posterior distributions of the parameters of the forward and counter jet afterglow emission. The results of $Model~(A)$ and $Model~(B)$ are shown in red and blue, respectively. The contours are at the $68\%$, $95\%$ and $99\%$ credible levels. The values are reported at the $68\%$ credible level.}
	\label{fig:posterior}
\end{figure*}

We consider a case that all of the physical parameters of the forward and counter jets are the same ($Model~(A)$), but find the counter jet radiation is unable to yield a detectable bump even by JWST (see Figure \ref{subfig:1a}). 
Note that in the afterglow modeling of GRBs, the microphysical parameters are found to be diverse for different events \citep[e.g.,][]{2002ApJ...571..779P}. This is also the case in the two component jet modeling of the afterglow emission of some GRBs such as GRB 051221A and GRB 080319B (i.e., the shock parameters are different for the narrow and wide components). If the microphysical parameters of the counter jet are higher than that of the forward jet, its radiation will be enhanced and hence may be able to yield a late time bump in the light curve. Motivated by such a prospect, in this work we also consider a ``general" case, in which the microphysical parameters of the counter jet are different from that of the forward jet.
In such a case, $\epsilon_{\rm e}$ and $\epsilon_{\rm B}$ for different jet components are independent free parameters (i.e., $Model~(B)$). All of the posterior results for $Model~(A)$ and $Model~(B)$ are presented in Table~\ref{tb:prior}. Accordingly, the best fitting light curves for these two models are shown in Figure~\ref{fig:bestfit} together with real observation data. To compare the parameter space of these two cases intuitively, the posterior distributions of the estimated parameters for the GRB afterglow and the kilonova afterglow are shown in Figure~\ref{fig:posterior} and Figure~\ref{fig:app}, respectively. The parameters (i.e., $E_0$, $\epsilon_{\rm e}$ and $\epsilon_{\rm B}$) are consistent with some recent studies on the afterglow of GRB 170817A \citep[e.g.,][]{2023ApJ...943...13W, 2023arXiv231002328R,2024arXiv240219359W}. Interestingly,  all these three works found a $\epsilon_{\rm e}\sim 0.0001$, which is much lower than the canonical value of $\sim 0.1$. At late times, because we include the additional contribution of the kilonova afterglow and use very late-time observations, the degeneracies among several parameters ($E_0$, $n_0$, $\epsilon_e$) are alleviated slightly compared with previous works \citep{2020ApJ...896..166R,2021ApJ...908..200W}.
By comparing the logarithm of the Bayesian evidence (ln(Z)) between $Model~(A)$ and $Model~(B)$, it turns out that both models can well fit the observations.

These best fitting light curves manifest that the forward jet afterglow dominates in the early stages, undergoing an initial rise and reaching a peak at around 160 days, followed by a rapid decline, as shown by the dashed lines in Figure~\ref{fig:bestfit}. 
At late times, the fluxes of the kilonova afterglow and the counter jet afterglow increase with possible detectability. Our results indicate that the kilonova and counter jet afterglow reaches their maximum fluxes at $\sim 600$ and $\sim 5000$ days. 
As shown in Figure~\ref{subfig:1a}, the best-fit counter jet afterglow reaches a lower peak flux than the kilonova afterglow, resulting in a nearly smooth decline in the overall afterglow light curves. This suggests that the component of the counter jet afterglow is hardly to be distinguished if its parameters are the same as the forward ones. In Figure~\ref{subfig:1b}, the peak fluxes of the counter jet afterglow is also lower than the fluxes of the kilonova afterglow and show a consistent profile compared with Figure~\ref{subfig:1a}.

At a confidence level of $68 \%$, the light curves of {\it Model (A)} exhibit a gradual decline even when the counter jet reaches its peak, while some residual radiation from the counter jet in {\it Model (B)} can reach relatively high peak values, as shown in Figure~\ref{subfig:1c} and Figure~\ref{subfig:1d}. The possible bump in the late time lightcurve is benefited from the chance that in {\it Model (B)} the microscopic parameters of the counter jet could be higher than that of the forward jet. Therefore, when considering the $68 \%$ confidence level of the light curves, it is possible to observe the counter jet radiation in $\it Model (B)$. 
The observability of the counter jet afterglow of GRB~170817A in multi-bands will be further discussed in the next section.

\section{Discussion}\label{sec:4}
Previous sections found out that the counter jet afterglow emission may be undetectable if the shock parameters are the same as that of the forward jet component (see Figure~\ref{subfig:1a}).  
However, the fit of the data leaving fit parameters to vary between the two jets can produce a brighter counter jet.
We also mention that the logarithm of the Bayesian evidence ln(Z) of $\it Model (B)$ minus ln(Z) of $\it Model (A)$ is nearly identical (see Figure~\ref{subfig:1b}). This is not a decisive evidence, yet. And it is not strong enough to give some preference to $\it Model (B)$. 
The key issue is that the blast wave driven by the counter jet has higher fractions of energy given to the electron acceleration as well as the magnetic energy than that of the forward jet. This can be understood as follows. Usually, the counter jet radiation becomes important as the blast wave gets decelerated to be non-relativistic. The deceleration timescale is expected to be $t_{\rm NR,cj} \approx 4\times 10^{3}~(1 + z)(E_{\rm cj,54} / n_{0})^{1/3}$ days, which is also roughly the peak radiation time \citep{2004ApJ...614L..17L}. Clearly, our numerical results are consistent with such an analytical estimate.   
In the non-relativistic phase, as long as the observer's frequency is below the so-called cooling frequency but above the typical synchrotron radiation frequency of the shocked electrons, the radiation flux is correlated with the shock parameters \citep[e.g.,][]{2004ApJ...614L..17L}, as
\begin{equation}\label{eq:11}
    F_{\nu} \propto \epsilon_{e}\epsilon_{B}^{3/4}.
\end{equation}
Note that this correlation holds for both the forward and counter jet components. Therefore, if the counter jet has driven a blast wave with substantially higher $\epsilon_{\rm e}$ and $\epsilon_{\rm B}$ than that of the forward jet, its synchrotron radiation will be able to dominate over other components (including the forward jet as well as the kilonova afterglow for the current event). 

Now we discuss the detection prospect of the counter jet radiation 
by several large-scale observational facilities, including the Hubble Space Telescope (HST), the Expanded Very Large Array (EVLA, \cite{2011ApJ...739L...1P}), the James Webb Space Telescope (JWST, \cite{2006SSRv..123..485G}), and the Extremely Large Telescope (ELT, \cite{2007Msngr.127...11G}). 
In Figure~\ref{subfig:1c} and Figure~\ref{subfig:1d}, the observation limits of HST, EVLA, JWST and ELT are calculated to be $\sim 2.1\times 10^{-2}~ \mu$Jy, $\sim 8.27~ \mu$Jy, $\sim 6.11\times 10^{-3}~ \mu$Jy, and $\sim 6.26\times 10^{-3}~ \mu$Jy in a 2-hour exposure \footnote{These limits for a detection are at the confidence level of $5\sigma$, which are calculated with the HST Exposure Time Calculator at \url{https://etc.stsci.edu}, the JWST Exposure Time Calculator at \url{https://jwst.etc.stsci.edu}, and the ELT Exposure Time Calculator at \url{https://www.eso.org/observing/etc}, respectively}, respectively.
By comparing the peak flux of the counter jet radiation at several wavelength with the sensitivities of these facilities, it is evident that for both the best fit and $68 \%$ confidence level scenarios, the counter jet radiation of {\it Model (A)} is too faint.
Similarly, in the best-fit scenario, the counter jet's peak radiation of $\it Model (B)$ is also significantly lower than the observation limits of these devices. Therefore, in these cases, the detection of the counter jet radiation is not promising.

However, in {\it Model (B)}, when we consider the $68 \%$ posterior distribution of the expected afterglow emission, the peak radiation of the counter jet may be above the observational limit of the instrument yielding a more promising detection prospect, as shown in Figure~\ref{subfig:1d}. Particularly in the near-infrared band (i.e., $\sim 8.4\times 10^{13}$ Hz), the peak radiations of the counter jet exceed the observational limits of JWST. In these observable regions, the afterglow emission of the counter jet dominates, as indicated by the dotted region in Figure~\ref{subfig:1d}. Therefore, if a plateau or a re-brightening appears in the late time afterglow light curve of GRB 170817A, it would serve as evidence for the presence of a counter jet.
In addition, the probability of detection of the counter jet afterglow in the near-infrared band by JWST would be higher than that of the other telescopes. Quantitatively, the probability of detecting the late-time ( $\sim 5000$ days) light curve bump by JWST in 2 hours is about 35\%. Though not very high, such a chance is non-negligible. At the peak time of the bump, the possibility is 39\% for the counter jet afterglow to dominate over other components, which accounts for all the detectability of the JWST.

\section{Conclusion}\label{sec:5}
As the first observed electromagnetic counterpart of the gravitational wave event GW170817, GRB 170817A took place at a luminosity distance of $d_{\rm L}\approx 40\, \rm Mpc$, which is one of the closet GRBs detected so far. Though GRB 170817A was subluminous, but it does mean that this burst was intrinsically weak. Instead, the low apparent isotropic energy of the prompt emission is simply due to the off-axis nature of the relativistic outflow.
The modeling of the forward jet radiation yields a high intrinsic kinetic energy. Given the short distance and the high intrinsic kinetic energy, GRB 170817A is an ideal target to search for the counter jet radiation.

In this work we have considered two scenarios: 1) the physical parameters of the counter jet and the forward jet are the same ({\it Model (A)}); 2) the $\epsilon_{\rm e}$ and $\epsilon_{\rm B}$ of the counter jet are free parameters ({\it Model (B)}). We have adopted these two models to conduct Bayesian fitting on the observed multi-wavelength afterglow data over a range of 9.2 to 1674 days and presented the corresponding fitting results in Figure~\ref{fig:bestfit} and Table~\ref{tb:prior}. 
It turns out that {\it Model (A)} yields a weak peak flux from the counter jet and lower than the kilonova afterglow at the same time and there is no prominent re-brightening of the afterglow emission at late times. 
{\it Model (B)} shares a similarity with {\it Model (A)} in terms of the absence of a clear peak radiation in its counter jet.
However, when we consider the fitting results at a $68 \%$ credible level, there is a chance that the peak afterglow emission from the counter jet in {\it Model (B)} could be significantly higher than the radiation from the forward jet at the same time (Figure~\ref{subfig:1d}). This means that the contribution from the counter jet's afterglow emission could be visible in very late time afterglow and has the potential to be observed by JWST in a few hours. 
The main reason for such an enhanced counter jet radiation is its relatively higher $\epsilon_{\rm e}$ and $\epsilon_{\rm B}$ than that of the forward jet. Though this still needs to be confirmed with the future observations, we would like to comment that diverse shock parameters for different jets have been found in GRB afterglow modeling and hence the assumptions made in {\it Model (B)} may be reasonable. 

Though the presence of a pair of jets from the GRB central engine has been widely accepted in this community, there is no solid observational evidence yet. GRB 170817A provides us an unprecedented chance to detect the counter jet radiation and hence prove the dual-sided jet scenario. We thus urge the continual observations in the next decade in particular in F356W band with JWST. The long term exposure with ELT/EVLA around 2030 would also be highly valuable.

\begin{acknowledgements}
We thank the anonymous referee for helpful comments and suggestions. We thank Yin-Jie Li, Shao-Peng Tang and Hao Wang for helpful discussions. We thank Lei Lei for estimating the detection limit of JWST.
This work is supported by the Natural Science Foundation of China (No. 11921003, No. 12233011, No. 11933010 and No. 12225305), and the New Cornerstone Science Foundation through the XPLORER PRIZE.

    $Software:$ {\tt Afterglowpy} (\citet{2020ApJ...896..166R}, \url{https://pypi.org/project/afterglowpy/}), {\tt Bilby} (\citet{2019ApJS..241...27A}, version 1.0.4, \url{https://git.ligo.org/lscsoft/bilby/}), {\tt Nessai} \citep{2021PhRvD.103j3006W}.
\end{acknowledgements}

\appendix
\section{The Posterior Distributions of the kilonova Afterglow Parameters}\label{sec:6}
\begin{figure*}[ht!]
	\centering
	\includegraphics[width=0.99\textwidth]{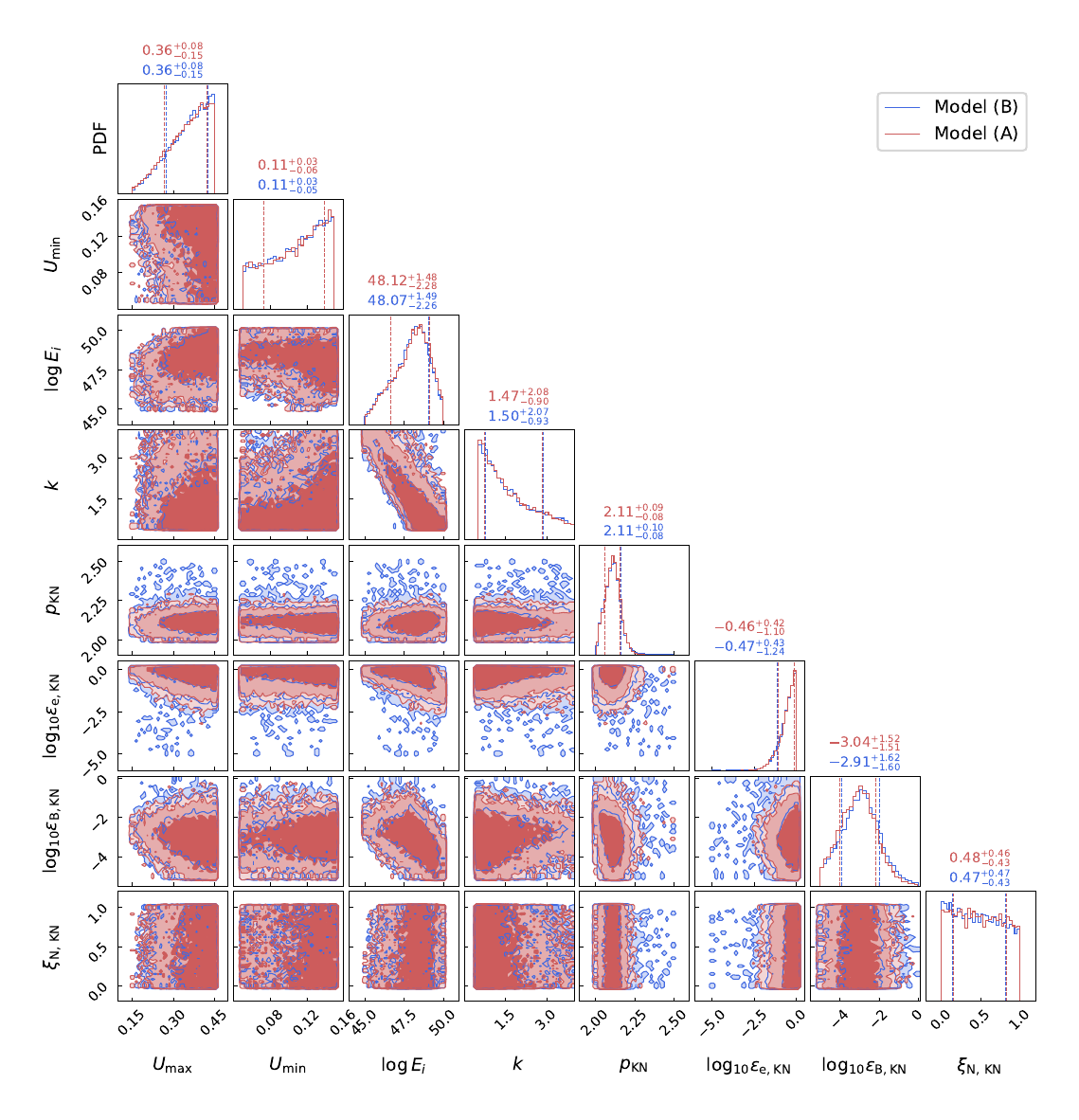}
    \caption{\small The posterior distributions of some parameters of kilonova afterglow modeling (see Figure.~\ref{fig:bestfit} for the reproduced light curves). The contours are at the $68\%$, $85\%$ and $90\%$ credible levels. The values are reported at the $68\%$ credible level.}
	\label{fig:app}
\end{figure*}

In Figure.~\ref{fig:app}, we present the posterior distributions of the kilonova afterglow modeling parameters (the best fitting results are shown in Figure.~\ref{fig:bestfit}). The spherical cocoon model roughly describes the evolution of kilonova afterglow, where the energy-velocity distribution follows a power-law distribution $E(U)=E_{\rm i} ({U}/{U_{\rm max}})^{-k}$, where $U$ is a dimensionless 4-velocity located in the $(U_{\rm min}, U_{\rm max})$ range, and $E_{\rm i}$ is the kinetic energy of the fastest material with $U=U_{\rm max}$. The mass ejected with velocity $U_{\rm max}$ is $M_{\rm ej} = {E_{\rm i}}/{((\gamma_{\rm max}-1)c^2)}$ \citep{2020ApJ...896..166R}, which is inferred to be $1.13^{+8.97}_{-1.08} \times 10^{-5}M_\odot$ with the current data.

\clearpage
\bibliography{References}
\bibliographystyle{aasjournal}
\end{document}